\documentclass{SCGE}
\usepackage{enumerate}
\usepackage{verbatim}
\begin{document}
\begin{picture}(0,0){\rm
\put(0,-20){\makebox[160truemm][l]{\bf {\sanhao\raisebox{2pt}{.}}
Article  {\sanhao\raisebox{1.5pt}{.}}}}}
\put(0,-34){\jiuwuhao {\textcolor[rgb]{0.5,0.5,0.5}{\sf 
}}}
\end{picture}
\def\bm{\boldsymbol}
\def\dl{\displaystyle}
\def\du{\end{document}}
\def\d{{\rm d}}
\def\e{{\rm e}}
\def\i{{\rm i}}
\def\pi{{\uppi}}
\Year{2016} %
\Month{??} %
\Vol{58} 
\No{?} %
\BeginPage{1} %
\AuthorMark{{\rm S.-X Yi \& S.-N. Zhang}}  
\AuthorMarkCite{{\rm S.-X Yi \& S.-N. Zhang} } 
\DOI{??} 
\ArtNo{??}
\title[Detecting super-Nyquist-frequency gravitational waves using a pulsar timing array]{Detecting super-Nyquist-frequency gravitational waves using a pulsar timing array}
\author[1,2]{Yi Shu-Xu}{}
\author[1*,3]{Zhang Shuang-Nan}{}
\address[{\rm1}]{Key Laboratory of Particle Astrophysics, Institute of High Energy Physics, Chinese Academy of Sciences, Beijing 100049, China;}
\address[{\rm2}]{University of Chinese Academy of Sciences, Beijing 100049, China;}
\address[{\rm3}]{Space Science Division, National Astronomical Observatories of
China, Chinese Academy of Sciences, Beijing 100012, China;}
\maketitle \vspace{-3.5mm}{\footnotesize\begin{center} Received ; accepted ; published
\end{center}}\vspace*{-5mm}
\begin{center}
\rule{16.5cm}{0.4pt}
\parbox{16.5cm}
{\begin{abstract} The maximum frequency of gravitational waves (GWs) detectable with traditional pulsar timing methods is set by the Nyquist frequency ($f_{\rm{Ny}}$) of the observation. Beyond this frequency, GWs leave no temporal-correlated signals; instead, they appear as white noise in the timing residuals. The variance of the GW-induced white noise is a function of the position of the pulsars relative to the GW source. By observing this unique functional form in the timing data, we propose that we can detect  GWs of  frequency $>$ $f_{\rm{Ny}}$ (super-Nyquist frequency GWs;  SNFGWs). We demonstrate the feasibility of the proposed method with simulated timing data. Using a selected dataset from the Parkes Pulsar Timing Array data release 1 and the North American Nanohertz Observatory for Gravitational Waves publicly available datasets, we try to detect the signals from single SNFGW sources. The result is consistent with no GW detection with 65.5\% probability. An all-sky map of the sensitivity of the selected pulsar timing array to single SNFGW sources is generated, and the position of the GW source where the selected pulsar timing array is most sensitive to is $\lambda_{\rm{s}}=-0.82$, $\beta_{\rm{s}}=-1.03$\,(rad); the corresponding minimum GW strain is $h=6.31\times10^{-11}$ at $f=1\times10^{-5}$\,Hz.
\end{abstract}}
\end{center}\vspace*{-0.6cm}
\begin{center}
\parbox{16.5cm}
{\bf\jiuhao gravitational wave, pulsar, black hole}
\end{center}
\begin{center}
{}
\CITA    
\end{center}
\textwidth=178truemm \textheight=236truemm
\wuhao\vspace*{1.5mm}
\begin{multicols}{2}
\renewcommand{\baselinestretch}{1.08} \baselineskip 12.2pt\parindent=10.8pt
\renewcommand{\thefootnote}
\section{Introduction}
\begin{figure}[H]
\centering
\includegraphics[width=7cm]{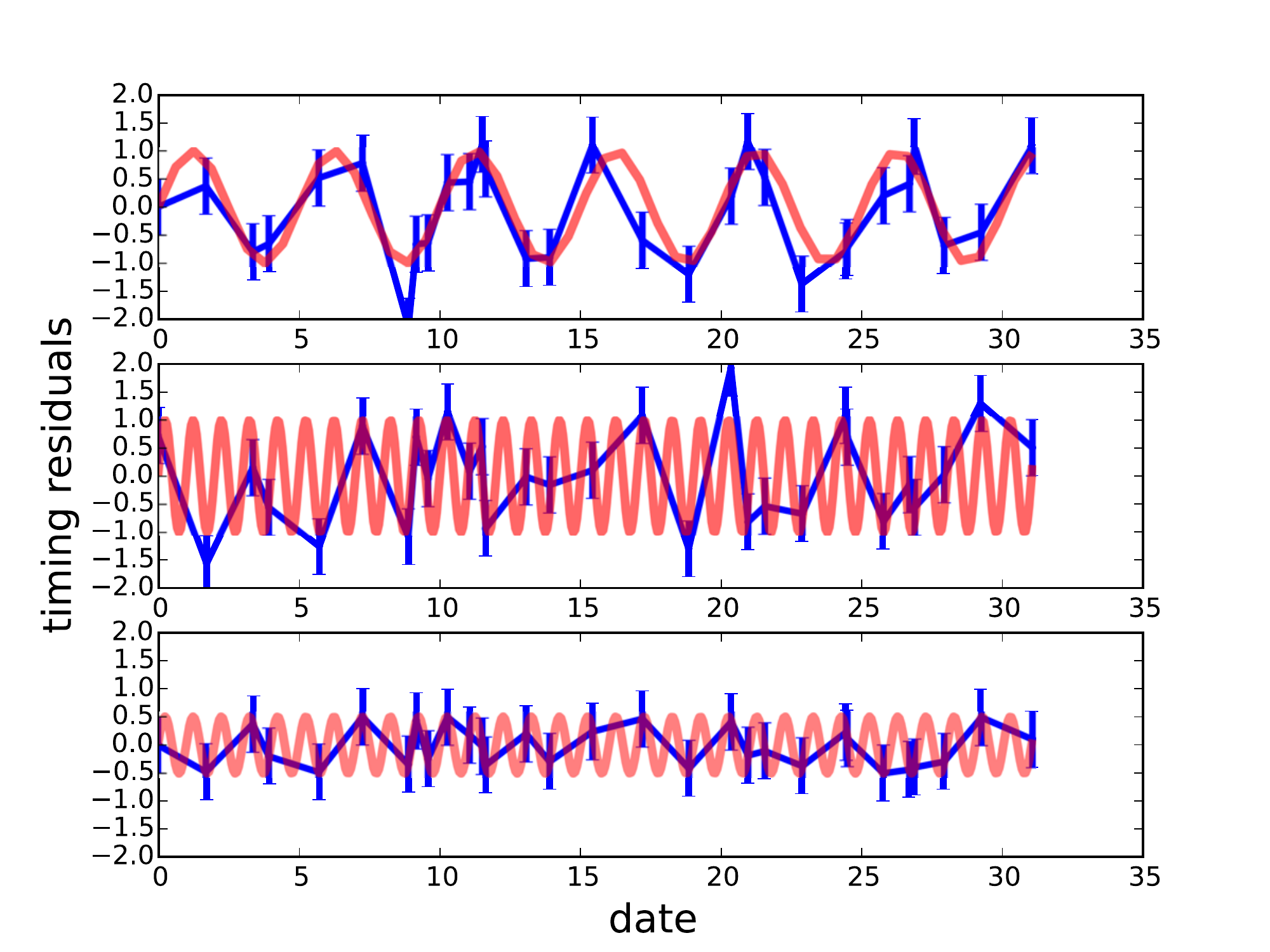}
\caption{\textbf{Upper panel:} When $f_{\rm {GW}}<f_{\rm{Ny}}$, the timing residuals (blue line) show a temporal-correlated structure following the wave form of the GW (red line). \textbf{Middle panel:} When $f_{\rm {GW}}>f_{\rm{Ny}}$, the timing residuals are white noise. \textbf{Bottom panel:} The weaker strain of SNFGWs leaves lower level of white noise in the timing residuals, compared to that in the middle panel.}\label{illustration}
\end{figure}

The recent direct detection of gravitational waves (GWs) \cite{2016PhRvL.116f1102A} marks the beginning of  GW astronomy era, after about five decades of effort on GW detection \cite{Weber, LIGO, LISA, IPTA, review, 2015SCPMA..58l5740L, 2015SCPMA..58l5738M, 2015SCPMA..58l5747B}. Among  the various proposed methods, the pulsar timing array (PTA) method shows promise in identifying GW-imprinted structure in the timing residuals of a number of pulsars \cite{Sazhin,Detweiler}. Although no detection has been made, more and more stringent upper limits of both individual GWs and GW background have been set using this method \cite{Nanobg,Nanosingle,EPTAbackground,PPTAsingle}.
The traditional pulsar timing method has an upper limit on the frequency of detectable GWs, which is known as the Nyquist frequency ($f_{\rm{Ny}}$). In the case of even sampling, $f_{\rm{Ny}}\sim N(2T_{\rm{obs}})^{-1}$, where $T_{\rm{obs}}$ is the observation time span, and $N$ is the number of the pulse time of arrival (TOA). In general, $f_{\rm{Ny}}$ is set by the sampling rate of the TOA. When $f_{\rm {GW}}>f_{\rm{Ny}}$, where $f_{\rm {GW}}$ is the frequency of GWs, GWs leave no temporal-correlated structures in the timing residuals of the pulsar and are therefore undetectable using the traditional pulsar timing method; instead, additional white noise will be left in the timing residuals, as the TOAs are not coherent in phase with the GWs (see the illustration in Figure \ref{illustration}). For the most typical biweekly observation scheme, the frequency upper bound is $\sim 10^{-7}$\, Hz. Although, in principle, GWs can be searched for at arbitrarily high frequency in the TOA using a Bayesian method \cite{2011PhDT.........2V}, the parameters of white timing noise will be completely correlated with the amplitude of the GWs, once $f_{\rm{Ny}}<f_{\rm {GW}}$, i.e., for super-Nyquist-frequency GWs (SNFGWs). However, an SNFGW will indicate itself by increasing the total white noise level in the timing residuals. The amplitude of additional GW-induced-white noise is a function of the coordinates of the pulsars relative to the GW source. In this paper, we propose that, by observing this unique relation between pulsar position and white noise variance, we can detect an SNFGW using a PTA.
The main purpose of this paper is to present the theory of the proposed method and also to demonstrate its feasibility with available data. The paper is organized as follows: The theory of our method is described at Section 2. We test the feasibility of our method on  simulated timing data in Section 3. A subset of pulsars is selected from the Parkes Pulsar Timing Array data release 1 (PPTA DR1) \cite{DR1} and the North American Nanohertz Observatory for Gravitational Waves  (NANOGrav dfg+12) publicly available datasets  \cite{Nanobg}. The selected PTA is used to search for SNFGWs using the proposed method in Section 4. In Section 5, we study the sensitivity of our method and the data to single SNFGW sources. We summarize our conclusions and discuss  potential targets and the advantages and shortcomings of the proposed method in the last section.
\section{Relation between the coordinates of pulsars and the SNFGW-induced timing noise}\label{theory}
\begin{figure}[H]
\centering
\includegraphics[width=7cm]{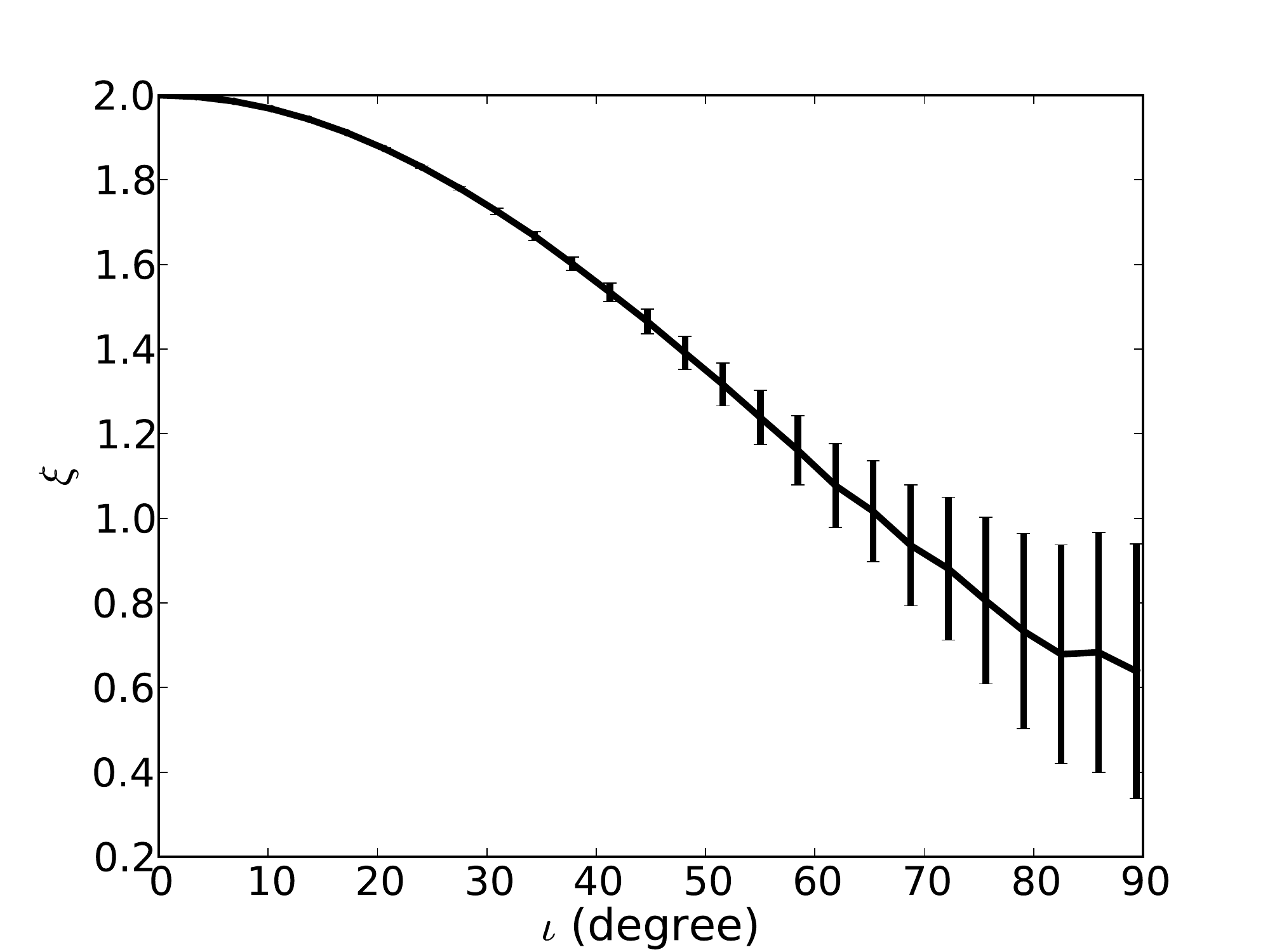}
\caption{Mean values of $\xi$ corresponding to 1,000 randomly distributed pulsars and random $\phi$ as a function of $\iota$. Error bars are the standard deviation of $\xi$.}\label{xifig}
\end{figure}

The total white timing noise consists of three parts: 1. white noise resulting from the intrinsic properties of pulses, e.g., single-pulse shape variability \cite{Shannon}, or from propagation through the interstellar medium (ISM) ($\sigma_{\rm{intrinsic}}$); 2. measurement uncertainties of the TOAs ($\sigma_{\rm{TOA}}$); and white noise induced by SNFGWs ($\sigma_{\rm{GW}}$). Since each $\sigma_{\rm{intrinsic}}$  is an intrinsic property of a pulsar, we expect that there is no correlation between $\sigma_{\rm{intrinsic}}$ from pulsar to pulsar.
Now we consider how the SNFGW determines $\sigma_{\rm{GW}}$ of a pulsar. Suppose that the right ascension and the declination of the pulsar are $\lambda$ and $\beta$, respectively, and that those of the GW source are $\lambda_{\rm{s}}$ and $\beta_{\rm{s}}$, respectively. The timing residuals brought by the GW are
\begin{equation}
r(t)=F_{+}A_{+}(t)+F_{\times}A_{\times}(t),\label{first}
\end{equation}
where $F_{+,\times}$ are the geometric factors \cite{LKJ11}
\begin{equation}
\begin{aligned}
F_{+}=&\frac{1}{4(1-\cos\theta)}
[(1+\sin^2\beta_{\rm{s}})\cos^2\beta\cos2(\lambda_{\rm{s}}-\lambda)\\
&-\sin2\beta_{\rm{s}}\sin2\beta\cos(\lambda_{\rm{s}}-\lambda)+\cos^2
\beta_{\rm{s}}(2-3\cos^2\beta)],\\
F_{\times}=&\frac{1}{2(1-\cos\theta)}
[\cos\beta_{\rm{s}}\sin2\beta\sin(\lambda_{\rm{s}}-\lambda)\\
&-\sin\beta_{\rm{s}}\cos^2\beta\sin2(\lambda_{\rm{s}}-\lambda)],\label{geo}
\end{aligned}
\end{equation}
where $\theta$ is the angle between the GW source and the pulsar, and $A_{+,\times}$ are
\begin{equation}
\begin{aligned}
A_{+}=&h/\omega[(1+\cos^2\iota)\cos2\phi\sin\omega t+2\cos\iota\sin2\phi\cos\omega t],\\
A_{\times}=&h/\omega[(1+\cos^2\iota)\sin2\phi\sin\omega t-2\cos\iota\cos2\phi\cos\omega t].\label{amplitute}
\end{aligned}
\end{equation}
In Equation (\ref{amplitute}), $h$ and $\omega$ are, respectively, the strain and the angular frequency of the GW, $\phi$ is the polarization angle of the GW, and $\iota$ is the inclination angle of the orbital plane of the GW source with respect to the line of sight. The initial phase of the GW is set to zero in Equation (\ref{amplitute}). We defer the treatment of the pulsar term to the discussion section; for now, we only consider the Earth term in Equation (\ref{amplitute}) for simplicity.
Equation (\ref{amplitute}) can be simplified as follows: We denote
 $K=h/\omega$, $\mu^2=F^2_{+}+F^2_{\times}$, and $\gamma=\arctan(F_{+}/F_{\times})+2\phi$; therefore, Equation (\ref{first}) can be rewritten as
\begin{equation}
r(t)=\mu K[(1+\cos^2\iota)\sin(\omega t)\sin\gamma-2\cos\iota\cos\omega t\cos\gamma].\label{four}
\end{equation}
We further denote
\begin{equation}
\xi^2=((1+\cos^2\iota)\sin\gamma)^2+(2\cos\iota\cos\gamma)^2\label{xieq}
\end{equation}
and
\begin{equation}
\psi=\arctan((1+\cos^2\iota)/2\cos\iota\tan\gamma),
\end{equation}
then Equation (\ref{four}) becomes
\begin{equation}
r(t)=\mu K\xi\sin(\omega t+\psi).\label{five}
\end{equation}
Since the sampling frequency of TOAs is less than the frequency of the sinusoid in Equation (\ref{five}), and the TOAs are random in the phase of the sinusoid, the resulting extra variance in the timing residuals is
 \begin{equation}
 \sigma^2_{\rm{GW}}=\frac{1}{2}\mu^2K^2\xi^2.\label{useful}
 \end{equation}
Among the three components of white noise, $\sigma^2_{\rm{intrinsic}}$ and $\sigma^2_{\rm{TOA}}$ are not correlated with the position of the pulsars, i.e., the parameter $\mu^2$; meanwhile, $\sigma^2_{\rm{GW}}$ is proportional to $\mu^2$. Therefore, as long as the strain of the GW is strong  enough to make $\sigma_{\rm{GW}}$ dominate, $\sigma^2\propto\mu^2$, where $\sigma$ is the rms of the timing residuals. Note that $\xi$ is in fact also a function of the positions of the pulsars; therefore, the correlation between $\sigma_{\rm{GW}}$ and $\mu^2$ will deviate from proportionality. $\xi$ as a function of $\iota$ is plotted in Figure \ref{xifig}: 1,000 pulsars were generated randomly in a  uniform distribution in the sky, and $\phi$ is uniformly random from 0$^\circ$ to $360^\circ$. The position of the GW source is assigned to $\lambda_{\rm {s}}=0, \beta_{\rm {s}}=0$. The corresponding $\xi$ is calculated using Equation (\ref{xieq}). In  Figure \ref{xifig}, the vertical axis indicates the mean values of $\xi$, and the error bars are the standard deviation of $\xi$.
When the inclination angle $\iota$ is far from 90$^\circ$, $\xi$ can be treated as a function of $\iota$, whereas when $\iota\sim90^\circ$, the scatter of $\xi$ around its mean value becomes more and more significant, which is the systematic scatter of Equation (\ref{useful}).
\section{Testing the method with simulated data}\label{simulate}
\begin{figure}[H]
\centering
\includegraphics[width=9 cm]{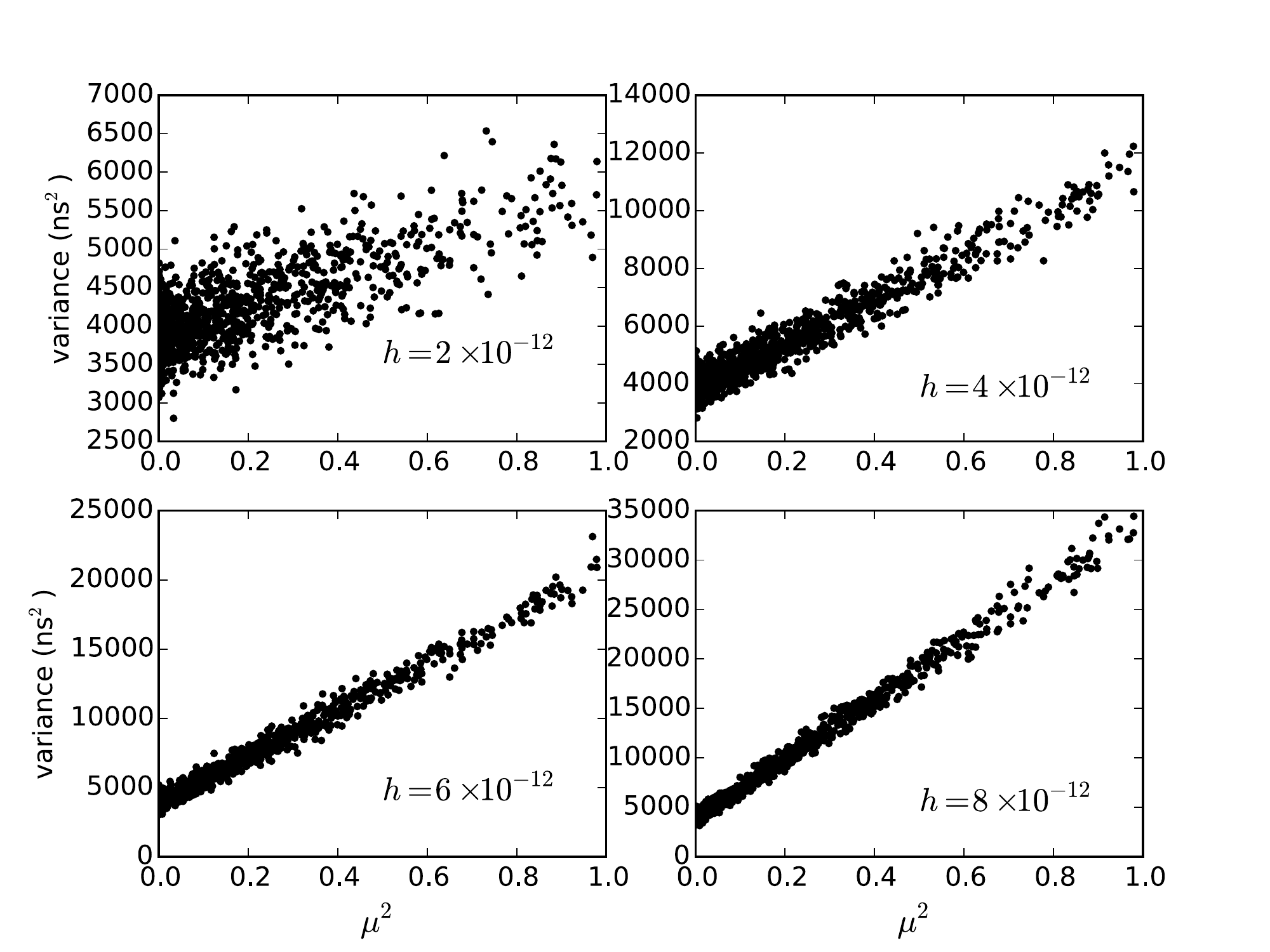}
\caption{Relationship between the timing residuals variance and $\mu^2$ of simulated timing residuals of pulsars with different injected GW strain $h$. The values of $h$ are indicated in each panel.}\label{figure1}
\end{figure}

\begin{figure}[H]
\centering
\includegraphics[width=9 cm]{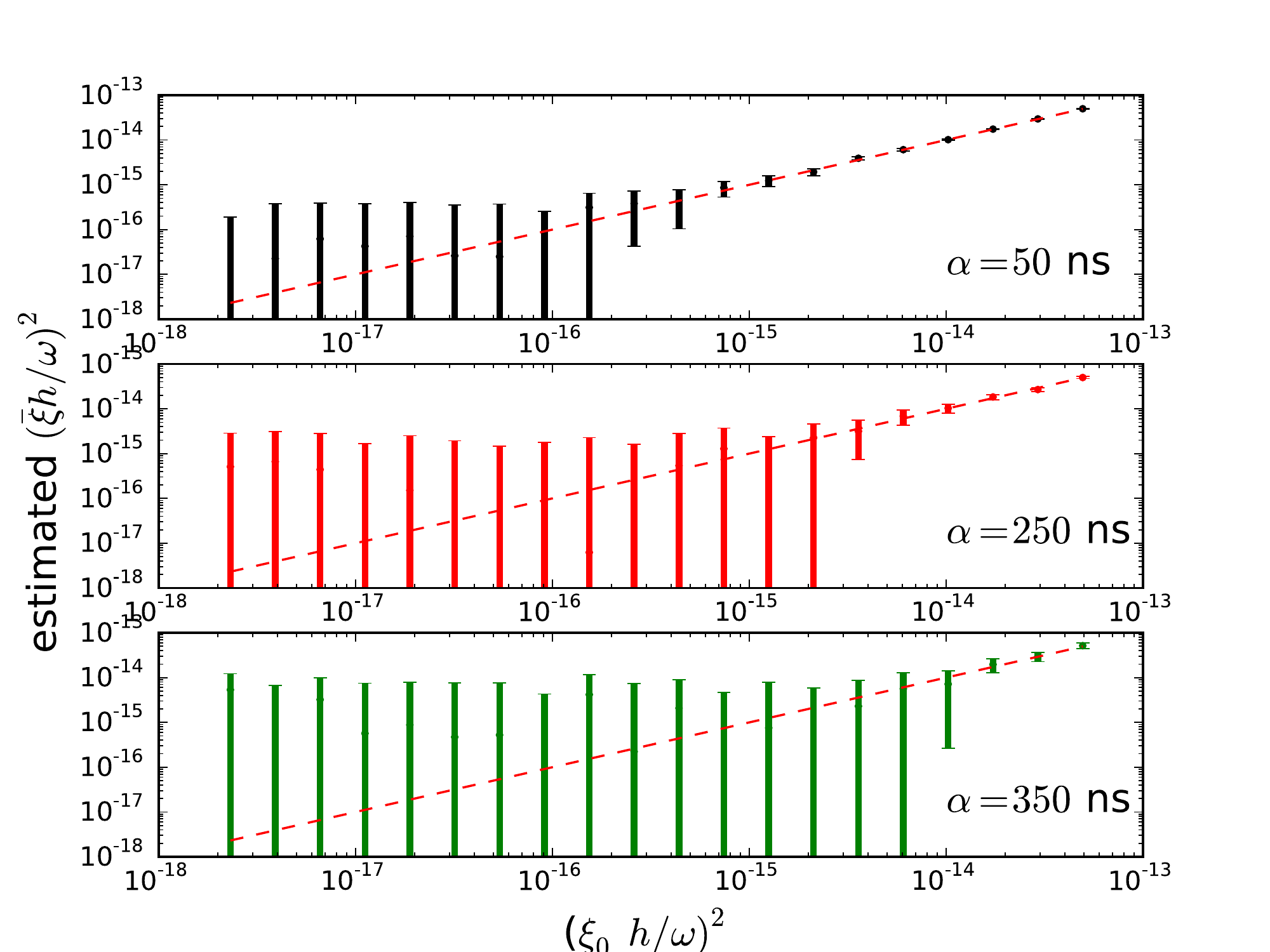}
\caption{Fitted slopes of the $\mu^2$-variance relationship, i.e., the estimated $(\bar{\xi}h/\omega)^2$ as functions of the injected $(\xi_0 h/\omega)^2$. Different intrinsic white noise levels $\alpha$ are indicated in each panel.}\label{figure2}
\end{figure}

To test the feasibility of the above-mentioned method, we generate simulated pulsar timing residuals for 1,000 pulsars. The pulsars are  uniformly distributed at random over the celestial sphere. The time spans, the total number of observations, the intrinsic white noise levels, and the TOA uncertainties are assigned according to the data of PSR J0437-4715 in PPTA DR1, while the dates of observations are randomly assigned for each pulsar.
The coordinates of the GW source are set at $\lambda_{\rm{s}}=0$, $\beta_{\rm{s}}=0$. Since the pulsars are uniformly distributed, the location of the GW source does not matter. The inclination angle of the source is set to $\iota=0.3$, and the polarization angle is  $\phi=0.1$, $f_{\rm {GW}}=1\times10^{-5}$ Hz, and $f_{\rm{Ny}}=1.58\times10^{-6}$ Hz; therefore, $f_{\rm {GW}}$ is super-Nyquist. The timing residual at each observation is assigned as
\begin{equation}
r(t_i)=G(0,\alpha)+G(0,E(t_i))+\mu K\xi\sin(2\pi ft_i),\label{seven}
\end{equation}
where $G(a,b)$ is a Gaussian with a  mean $a$ and a standard deviation $b$; $\alpha$ is the intrinsic white noise level, which is set to $\alpha=50$ ns; $E(t_i)$ is the TOA uncertainty at each $t_i$. Figure \ref{figure1}  shows the relationship between the variance of timing residuals and the position parameter $\mu^2$ of the simulated timing data of the pulsars. The correlation becomes more and more significant with increasing GW strain.
The slope of the above-mentioned $\mu^2$-variance relationship is equal to the combination $\bar{\xi}h/\omega$ of the GW, where the factor $\bar{\xi}$ is included to take account of $\xi$ for different pulsars.  We inject different  GW strain into the timing residuals and perform linear fit to the resulting $\mu^2$-variance pairs, and we plot the fitted slope, i.e., the estimated $(\bar{\xi}h/\omega)^2$ as a function of the injected value of $(\xi_0 h/\omega)^2,$ in Figure \ref{figure2}, where $\xi_0$ is the parameter $\xi$ when $\gamma$ is fixed to $2\phi$ in  Equation (\ref{four}). The error bars show the $3\sigma$ error of the slope given by the fitting process. When the injected GW strain is small, the fitted slopes have large uncertainties and deviate from the injected values. As the injected GW strain increases  enough, the slopes fall onto the red dashed lines, where the estimated $(\bar{\xi}h/\omega)^2$ equals the injected $(\xi_0 h/\omega)^2$. By changing the intrinsic white noise level $\alpha$, we find that higher intrinsic white noise decreases the sensitivity of the PTA to the GW; this conclusion is in accordance with  traditional pulsar timing methods.
\section{Detecting SNFGWs using the selected PTA from PPTA DR1 and NANOG\MakeLowercase{rav dfg+12}}
\begin{figure*}
\centering
\includegraphics[width=0.8\textwidth]{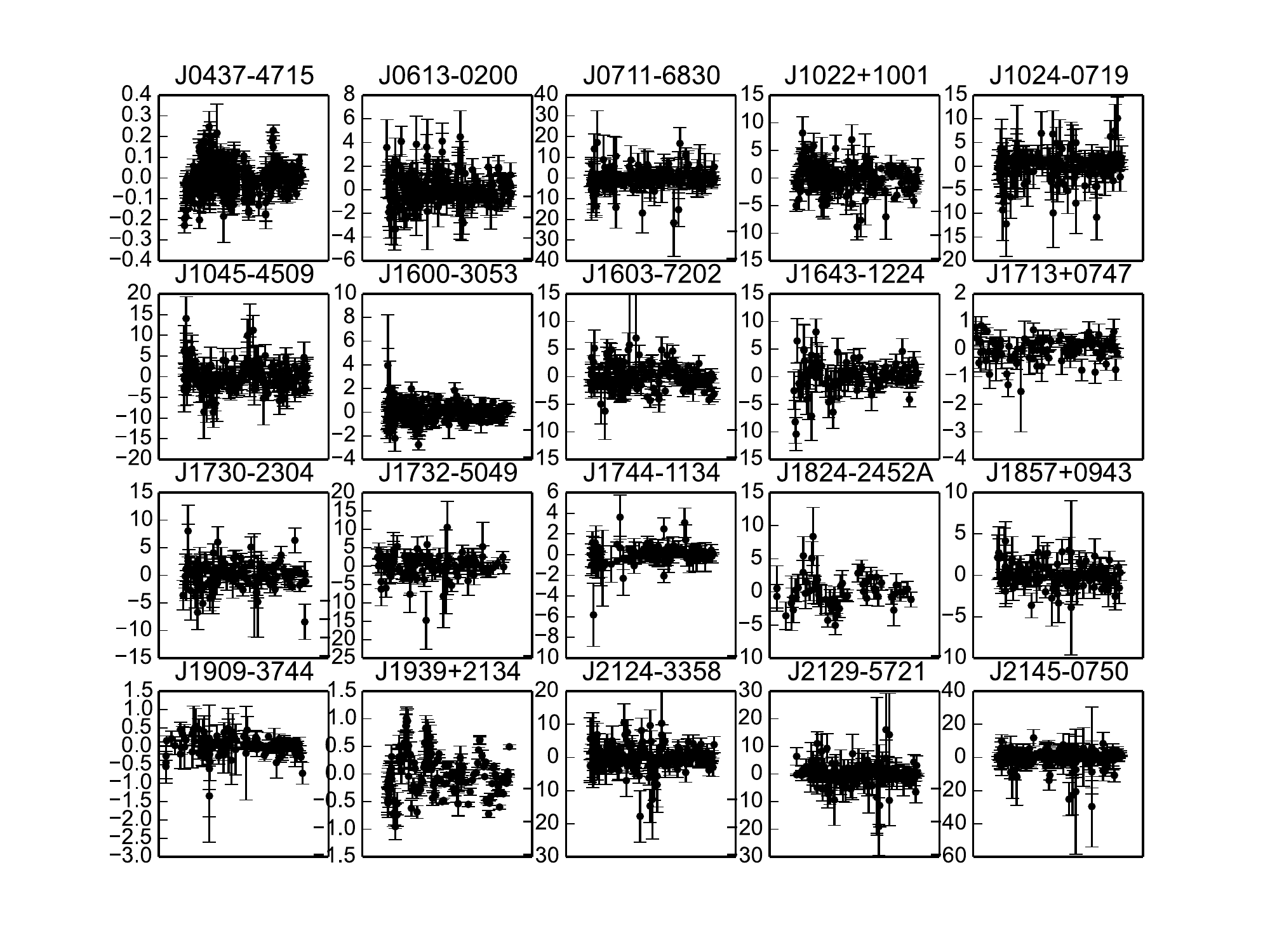}
\caption{Timing residuals of PPTA DR1 pulsars. The names of the pulsars are indicated at the top of each panel. The axes are hidden for clarity. }\label{timingresiduals}
\end{figure*}

\begin{figure*}
\centering
\includegraphics[width=0.8\textwidth]{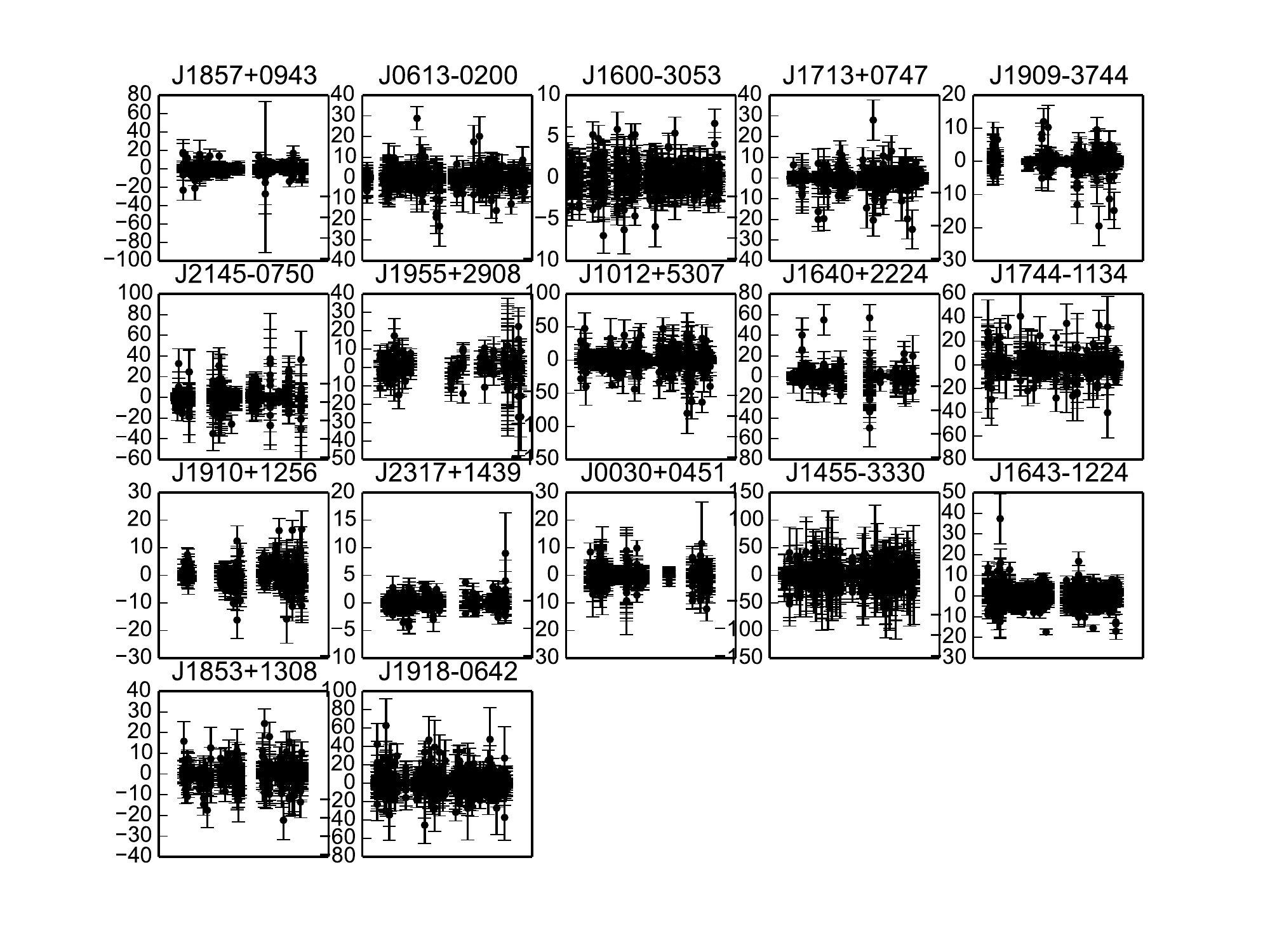}
\caption{Timing residuals of NANOGrav dfg+12 pulsars. The names of the pulsars are indicated at the top of each panel. The axes are hidden for clarity.}\label{timingresiduals2}
\end{figure*}

\begin{figure*}
\centering
\includegraphics[width=10 cm]{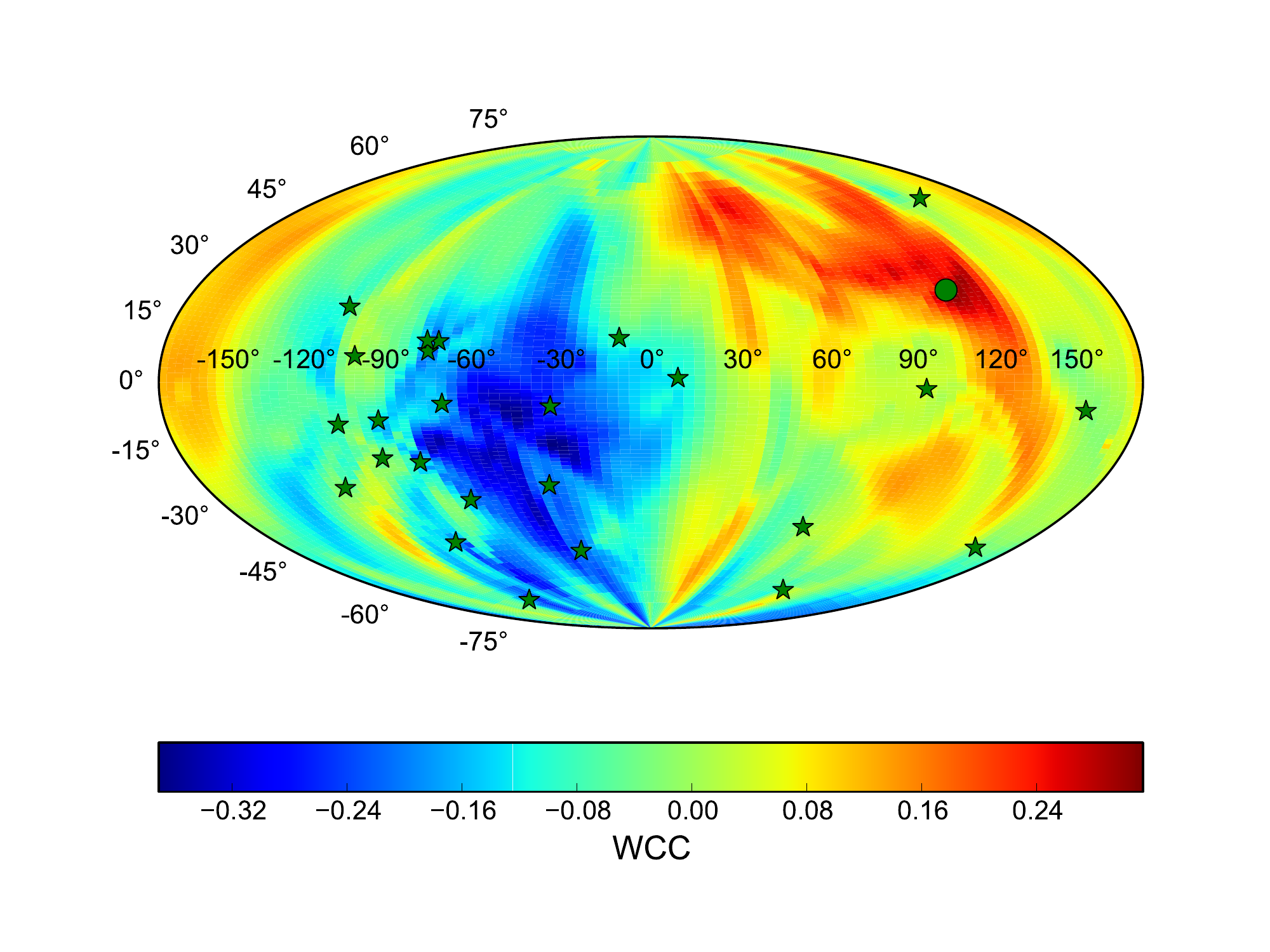}
\caption{All-sky map of the WCC between $\log\mu^2$ and $\log\sigma^2_{\rm{remain}}$. The green stars indicate the locations of the pulsars used; the green circle is the position where the WCC reaches its maximum.}\label{skymap}
\end{figure*}

\begin{figure}[H]
\centering
\includegraphics[width=9 cm]{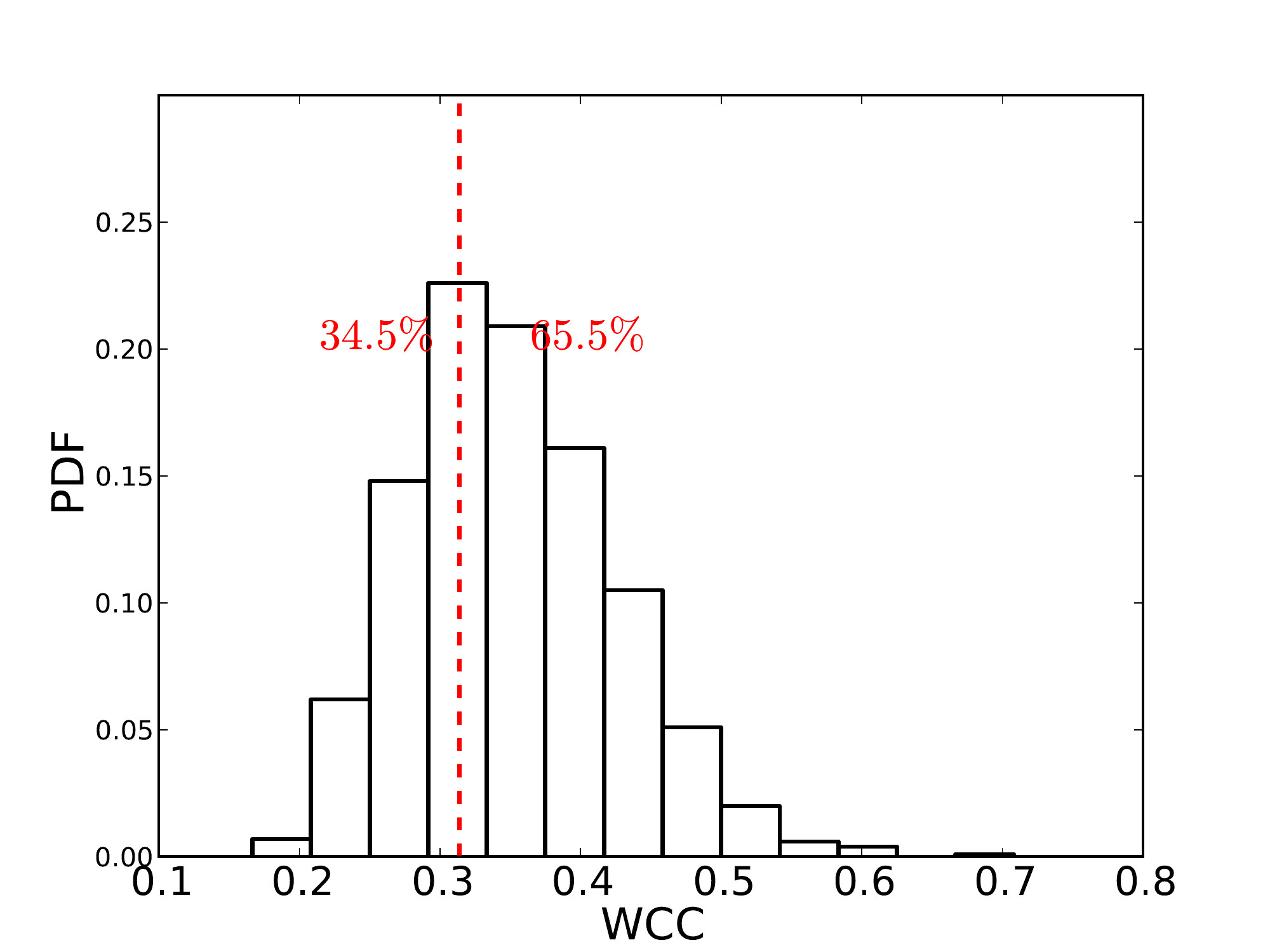}
\caption{Normalized probability density distribution (PDF) of the maximum WCC between $\log\mu^2$-$\log\sigma_{\rm{remain}}^2$ of 1,000 permutations of $\sigma_{\rm{remain}}$. The vertical dashed line indicates the observed WCC 0.31; the probability that any permutation has the maximum WCC larger than the observed one is 65.5\%. Therefore, the probability that the observed $\log\mu^2$-$\log\sigma_{\rm{remain}}^2$ correlation is the consequence of intrinsic white noise of the pulsars is 65.5\%, and this result indicates a nondetection.}\label{histo}
\end{figure}

In this section, we demonstrate how the proposed method can be used with real pulsar timing data by replacing the simulated pulsar timing data with  real data selected from PPTA DR1 and NANOGrav dfg+12. As mentioned above, $\sigma_{\rm{TOA}}$, $\sigma_{\rm{intrinsic}}$ and $\sigma_{\rm{GW}}$ all contribute to the total residuals budget. The contribution of $\sigma_{\rm{TOA}}$ can be calculated from the uncertainty of the TOA and can be removed from the total residuals rms. We denote the rms of the remaining residuals as $\sigma_{\rm{remain}}$. $\sigma_{\rm{remain}}$ is composed of $\sigma_{\rm{intrinsic}}$ and $\sigma_{\rm{GW}}$. In the ideal case, $\sigma_{\rm{intrinsic}}$ values of these pulsars are intrinsic to the individual pulsars and they do not correlate with the PTA. Therefore the only component that makes $\sigma_{\rm{remain}}$ correlate with the PTA is $\sigma_{\rm{GW}}$. However, in practice, owing to the complexity of the observing systems (i.e., front end/back end combinations), the apparent TOA uncertainty cannot faithfully reflect $\sigma_{\rm{TOA}}$. As a result, the rms of some instrument-related timing residuals enters $\sigma_{\rm{remain}}$, which are also correlated in complicated ways but beyond the scope of this current work. Therefore, the estimated significance of GW detection in this section should be considered only as an upper limit or optimistic evaluation.
The timing residuals are obtained by fitting the TOA using \texttt{TEMPO2} \cite{tempo2} and the variance of the timing residuals are thus calculated. The timing residuals of PSR J1939+2134 and J1824-2452A are polynomial whitened, while the data of other pulsars are fitted using the downloaded ephemeris. The bias and scaling factors EQUAD and EFAC are all set to zero in the fitting. The resulting timing residuals of pulsars of PPTA DR1 and NANOGrav dfg+12 are plotted in Figures \ref{timingresiduals} and \ref{timingresiduals2}, respectively. $\sigma_{\rm{remain}}$ of each pulsar is estimated as follows:
We generate a new series of timing residuals for each pulsar such that
\begin{equation}
r_{\rm{sim}}(t_i)=G(0,\sigma_{\rm{remain}})+G(0,E(t_i))
\end{equation}
from a small starting value; we increase $\sigma_{\rm{remain}}$ until the variance of the simulated timing residuals (Var$_{\rm{sim}}$) equals  the real variance (Var$_{\rm{real}}$). In practice, we consider these two quantities to be identical when the relative difference ($|\rm{Var_{sim}}-\rm{Var_{real}}|/\rm{Var_{real}})<10\%$. We list the total rms ($\sigma_{\rm{total}}$), $\sigma_{\rm{remain}}$, and the average of the TOA uncertainties in Tables \ref{table} and \ref{table2}.
\begin{table}[H]
\scriptsize
\begin{tabular}{lrrr}
\hline
name & $\sigma_{\rm{total}}$ (ns) & $\sigma_{\rm{remain}}$ (ns) & Ave. $\Delta$TOA (ns)\\
\hline
\hline
J0437-4715 &69&42&41\\
J0613-0200 &1301&19&1042\\
J0711-6830 &4405&3&3357\\
J1022+1001 &2315&625&1327\\
J1024-0719 &2981&7&2279\\
J1045-4509 &3230&15&2596\\
J1600-3053 &758&9&540\\
J1603-7202 &2207&24&1283\\
J1643-1224 &2722&3&2022\\
J1713+0747 &424&9&269\\
J1730-2304 &2296&4&1677\\
J1732-5049 &3224&2&2585\\
J1744-1134 &920&6&573\\
J1824-2452A &2337&13&1687\\
J1857+0943 &1384&8&1292\\
J1909-3744 &255&8&232\\
J1939+2134 &402&295&142\\
J2124-3358 &3641&8&2602\\
J2129-5721 &3703&5&3017\\
J2145-0750 &3532&3&2175\\
\hline
\end{tabular}
\caption{$\sigma_{\rm{total}}$, $\sigma_{\rm{remain}}$ and the average of TOA uncertainties (Ave. $\Delta$TOA) of PPTA DR1; the unit is nanosecond.}\label{table}
\end{table}
\begin{table}[H]
\scriptsize
\begin{tabular}{lrrr}
\hline
name & $\sigma_{\rm{total}}$ (ns) & $\sigma_{\rm{remain}}$ (ns) & Ave. $\Delta$TOA (ns)\\
\hline
\hline
J1857+0943 &3492&11&2050\\
J0613-0200 &2884&3&2058\\
J1600-3053 &1549&87&1303\\
J1713+0747 &1842&143&817\\
J1909-3744 &1654&317&811\\
J2145-0750 &7173&8&4836\\
J1955+2908 &6104&5001&7107\\
J1012+5307 &7493&11&4457\\
J1640+2224 &6353&10&2753\\
J1744-1134 &4530&2&2341\\
J1910+1256 &3308&17&2233\\
J2317+1439 &1031&31&629\\
J0030+0451 &2646&3&2080\\
J1455-3330 &12747&5001&13868\\
J1643-1224 &3571&2039&2243\\
J1853+1308 &4208&2&3431\\
J1918-0642 &7172&9&5180\\
\hline
\end{tabular}
\caption{$\sigma_{\rm{total}}$, $\sigma_{\rm{remain}}$ and the average of TOA uncertainties (Ave. $\Delta$TOA) of NANOGrav dfg+12; the unit is nanosecond.}\label{table2}
\end{table}
From Tables \ref{table} and \ref{table2} we chose  pulsars with $\sigma_{\rm{remain}}<100$  ns, and, for pulsars shared by both tables, the smaller $\sigma_{\rm{remain}}$ values are selected. The resulting pulsars are listed in Table \ref{table3}.
\begin{table}[H]
\scriptsize
\begin{tabular}{lr|lr}
\hline
name & $\sigma_{\rm{remain}}$ (ns) & name & $\sigma_{\rm{remain}}$ (ns)\\
\hline
\hline
J0437-4715 & 42 & J1857+0943 & 8 \\
J0613-0200 & 19 & J1909-3744 & 8\\
J0711-6830 & 3 & J2124-3358 & 8\\
J1024-0719 & 7 & J2129-5721 & 5\\
J1045-4509 & 15 & J2145-0750 & 3\\
J1600-3053 & 9 & J1012+5307 & 11\\
J1603-7202 & 24 & J1640+2224 & 10\\
J1643-1224 & 3 & J1910+1256 & 17\\
J1713+0747 & 9 & J2317+1439 & 31\\
J1730-2304 & 4 & J0030+0451 & 3\\
J1732-5049 & 2 & J1853+1308 & 2\\
J1744-1134 & 6 & J1918-0642 & 9\\
J1824-2452A & 13 & & \\
\hline
\end{tabular}
\caption{Selected 25 pulsars in this work.}\label{table3}
\end{table}
After $\sigma_{\rm{remain},\textit{i}}$ is known for the $i$th pulsar (where $i$ ranges from 1 to $N_{\rm{psr}}$, where $N_{\rm{psr}}$ is the number of pulsars in the PTA), we need to obtain $\mu^2_i$ using Equation (\ref{geo})  to test the correlation described in Equation (\ref{useful}). Since the location of the GW source is unknown, we divide the celestial sphere into $100\times100$ equal-area grids. For each grid we suppose that the GW source is locates within it and we calculate $\mu^2_i$. We want to test the correlations between $\mu^2_i$ and $\sigma_{\rm{remain},\textit{i}}^2$. Owing to the nature of Pearson correlation coefficient (PCC),  data with greater distance to the barycenter contribute more to the PCC. As a result, if we use the PCC to study the correlation,  the minority of the pulsars that have the largest $\sigma_{\rm{remain}}$ will dominate the PCC of $\mu^2_i$-$\sigma_{\rm{remain},\textit{i}}^2$. To avoid this problem, we study the correlation on a logarithm scale, in which the scatter of data decreases. In this way, we also assign less weight to the outliers. We therefore use the weighted correlation coefficient (WCC) between $\log\mu^2_i$ and $\log\sigma_{\rm{remain},\textit{i}}^2$.
The WCC between two lists of data $X_i$ and $Y_i$ is defined as \begin{equation}
        r_w=\frac{\sum w_iX_iY_i-\sum w_iX_i\sum w_iY_i}{\sqrt{\sum w_iX_i^2-(\sum w_iX_i)^2}\sqrt{\sum w_iY_i^2-(\sum w_iY_i)^2}},\end{equation}
        where  $w_i$ are the normalized weights. The weight that we assign to each data point is its distance to the barycenter.
A loop of all the sky grids gives the all-sky map of  $\log\mu^2_i$-$\log\sigma_{\rm{remain},\textit{i}}^2$-WCC, which is plotted in Figure \ref{skymap}; the color scale indicates the WCC. The celestial grid where $\log\mu^2_i$-$\log\sigma_{\rm{remain},\textit{i}}^2$-WCC is maximum is indicated by the green circle. The coordinates of this point are $\lambda_{\rm{s}}=1.95$, $\beta_{\rm{s}}=0.48$ (rad), and the corresponding log-log-WCC is 0.31.
We need to know the probability that the above $\log\mu^2$-$\log\sigma_{\rm{remain}}^2$ correlation is due to the intrinsic white noise of the PTA. Therefore, we randomly shuffle the $\sigma_{\rm{remain},\textit{i}}$ values of the pulsars  1,000 times, and we calculate the all-sky map of log-log-WCC and the largest WCC for each permutation. We then get the distribution of the maximum WCC, which is plotted in Figure \ref{histo}. We notice that 65.5\% of the realizations have a maximum WCC larger than the observed value 0.31; thus the probability that the observed $\log\mu^2$-$\log\sigma_{\rm{remain}}^2$ correlation is the consequence of intrinsic white noise of the pulsars is 65.5\%, and the result is consistent with a nondetection.
\section{Sensitivity to single SNFGW sources}
In the above section we found a nondetection result and we want to study the sensitivity of this method to single SNFGW sources using PPTA DR1 and NANOGrav dfg+12 data. The procedure is outlined as follows:
\begin{enumerate}[1)]
\item{Divide the sky sphere uniformly into $100\times100$ grids. In each grid, put an SNFGW source. The inclination angle is set to optimal $\iota=0$, and the frequency of the GW is $f=1\times10^{-5}$ Hz.}
\item{Starting from a small GW strain $h$ value and a random polarization angle $\phi$, generate a series of timing residuals based on Equation (\ref{seven}).}
\item{Follow the SNFGW source-detecting procedure described in the above section. Increase $h$ and return  to step 2, until the detection significance reaches 99\%.}
\item{Record the current value of $h$ as the minimum GW strain that the dataset is sensitive to. Move to the next grid point of the sky sphere.}
\end{enumerate}
\begin{figure*}
\centering
\includegraphics[width=10 cm]{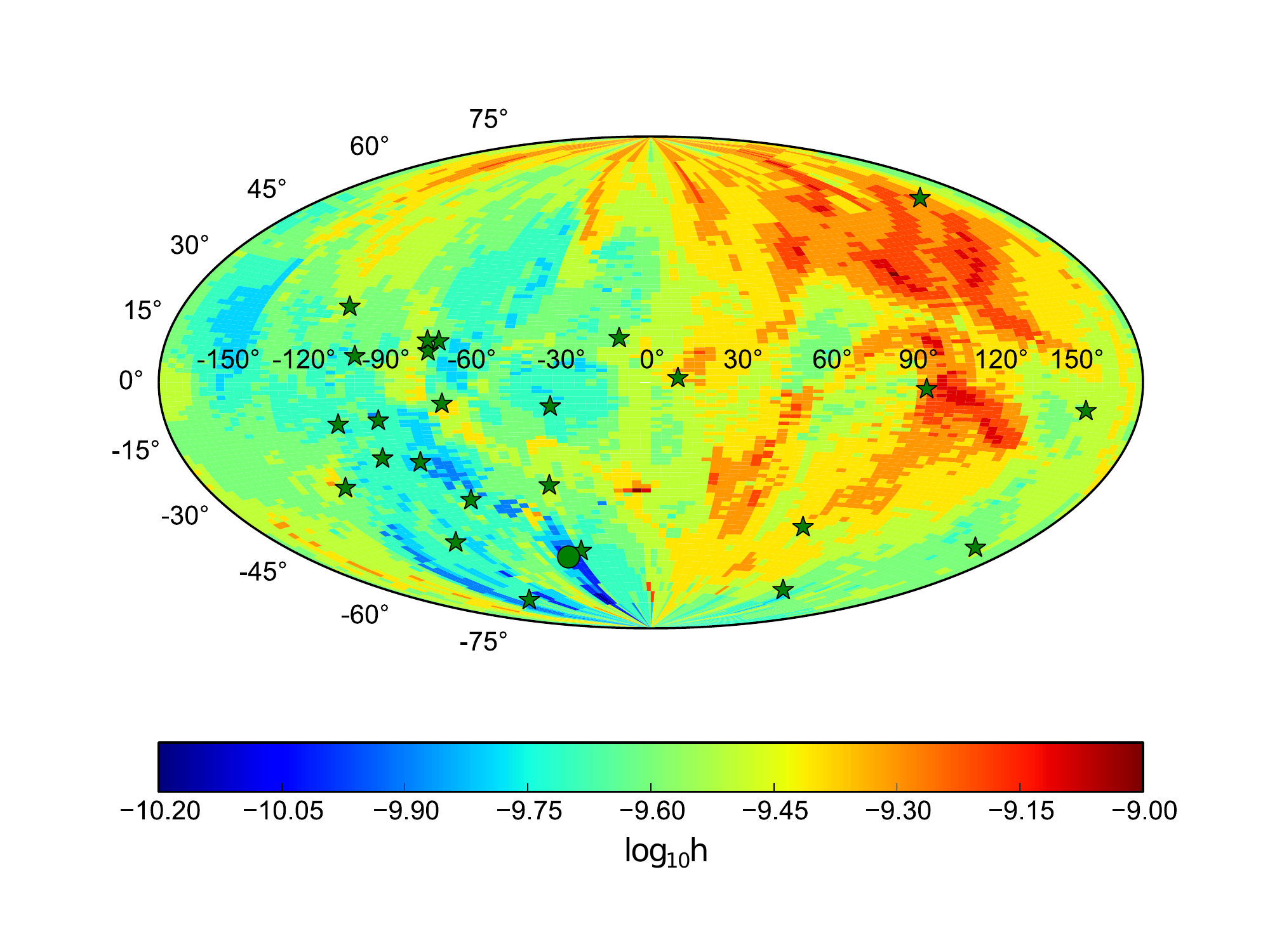}
\caption{All-sky map of the sensitivity of our data to single SNFGW sources. The green stars indicate the locations of the pulsars used; the green circle is the position where our PTA is most sensitive to the GW source.}\label{sensmap}
\end{figure*}
The resulting all-sky map of sensitivity is presented in Figure \ref{sensmap}. The position of the GW source where the selected PTA is most sensitive to is $\lambda_{\rm{s}}=-0.82$, $\beta_{\rm{s}}=-1.03$ (rad), which is indicated with a green circle in Figure \ref{sensmap}; the corresponding minimum is $h=6.31\times10^{-11}$ at $f=1\times10^{-5}$ Hz. According to Equation (\ref{amplitute}), the sensitive $h$  scales with the frequency $f$. We present our sensitivity results in the super-Nyquist band in Figure \ref{curve}, compared with limits from previous PTA works (the sensitivity curves of the LIGO and the proposed LISA).
\begin{figure}[H]
\centering
\includegraphics[width=6.5 cm, angle=-90]{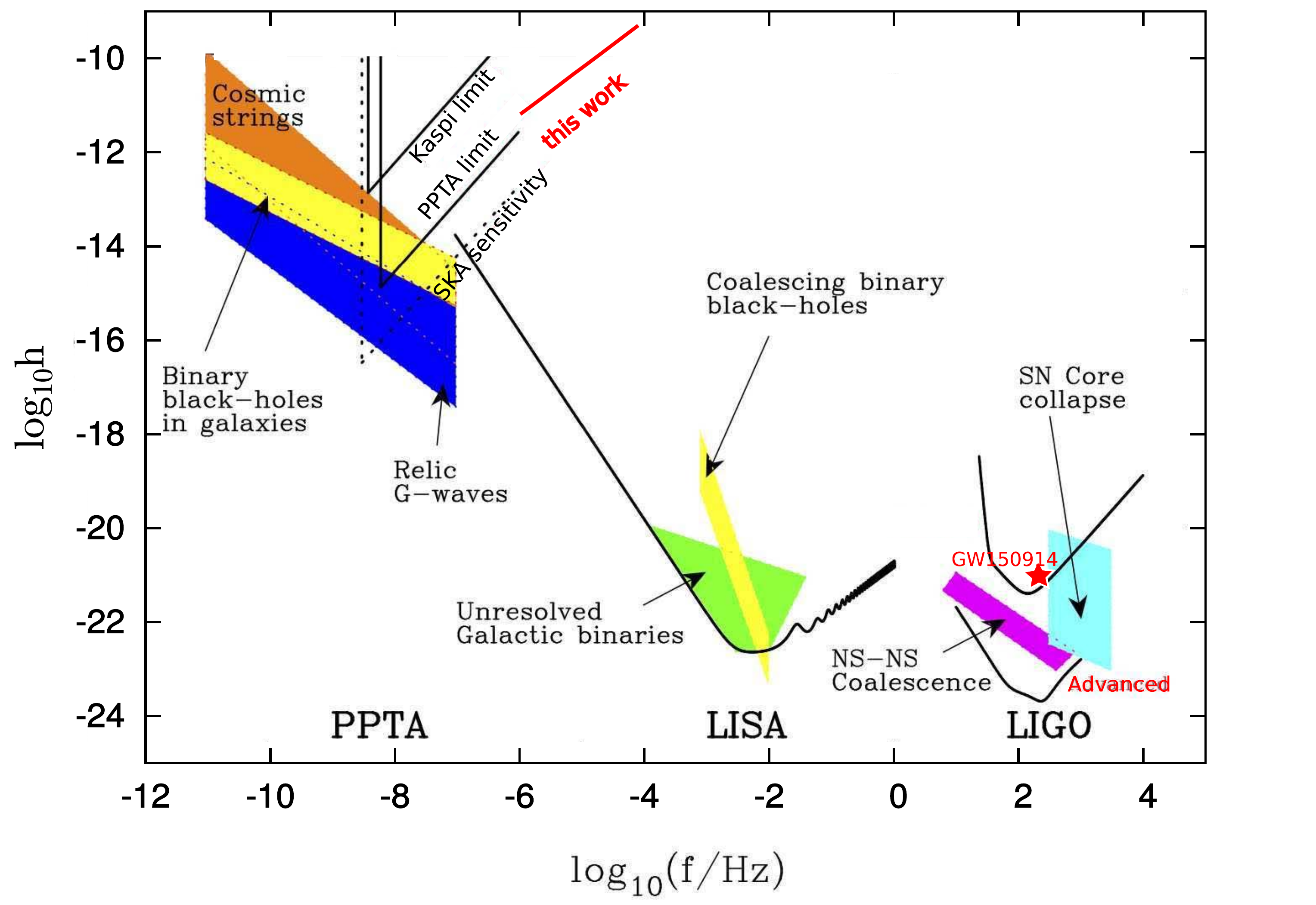}
\caption{The sensitivity of the method described in this work to single gravitational wave sources for the optimal binary orientation, compared with the sensitivity or upper limits of other methods. The red star marks the peak signal of the first GW event GW150914 \cite{2016PhRvL.116f1102A}. The figure is edited from \cite{Curves}.
}\label{curve}
\end{figure}
\section{Summary, Conclusions, and Discussion}
\subsection{Summary}
\begin{enumerate}
\item{The theory of the method for detecting SNFGWs is presented in Section \ref{theory}.}
\item{We tested the theory in Section \ref{simulate} using simulated PTA data and showed that studying the correlation between $\sigma^2_{\rm{GW}}$ and $\mu^2$ can serve as a method for detecting SNFGWs. }
\item{The all-sky map of the WCC between $\log\mu^2$ and $\log\sigma^2_{\rm{remain}}$ is shown in Figure \ref{skymap}.}
\item{The all-sky map of the sensitivity of the selected PTA to single SNFGW sources is shown in Figure \ref{sensmap}.}
\end{enumerate}
\subsection{Conclusions}
We summarize our conclusions as follows:
\textbf{Theory:}
\begin{enumerate}
\item{SNFGWs leave additional white noise in the timing residuals.}
\item{$\sigma^2_{\rm{GW}}$ of the GW-induced white noise is proportional to the parameter $\mu^2$, and the $\sigma^2_{\rm{GW}}$-$\mu^2$ proportional relationship is scaled by $(1/2)\xi^2(h/\omega)^2$. 
  For definitions of the other symbols see Section \ref{theory}.
}
\end{enumerate}
\textbf{Simulation:}
\begin{enumerate}
\item{When $f_{\rm {GW}}$ is given, the stronger GW strain will give a more significant $\sigma^2_{\rm{GW}}$-$\mu^2$ correlation.}
\item{The combination $(\xi_0h/\omega)^2$ of the GW source can be estimated by fitting the $\sigma^2_{\rm{GW}}$-$\mu^2$ relationship.}
\item{The GW strain needs to be higher than a lower limit to make the $\sigma^2_{\rm{GW}}$-$\mu^2$ correlation unambiguous. The lower limit of GW strain increases with the intrinsic white noise level of the PTA.}
\end{enumerate}
\textbf{Detection:}
\begin{enumerate}
\item{The coordinates of the GW source where the WCC is optimized are $\lambda_{\rm{s}}=1.95$, $\beta_{\rm{s}}=0.48$ (rad), and the corresponding WCC is 0.31.}
\item{Monte Carlo simulation indicates that the probability that the observed $\log\mu^2$-$\log\sigma_{\rm{remain}}^2$ correlation is the consequence of the intrinsic white noise of the pulsars is 65.5\%.
}
\end{enumerate}
\textbf{Sensitivity to single SNFGW  sources:}
\begin{enumerate}
\item{The position of the GW source where the selected PTA is most sensitive to is $\lambda_{\rm{s}}=-0.82$, $\beta_{\rm{s}}=-1.03$ (rad); the corresponding minimum GW strain is $h=6.31\times10^{-11}$ at $f=1\times10^{-5}$ Hz.}
\end{enumerate}
\subsection{Discussion}
\subsubsection{Target sources}
The SNFGW sources that we aim to study are the merging supermassive black hole binaries (SMBHBs). The frequency of the GW is higher than the typical $f_{\rm{Ny}}$ of a PTA, i.e., $\sim10^{-7}$ Hz. Therefore, we can use filtering techniques to remove red noise from other origins. We treat the strain and $f_{\rm {GW}}$ as steady throughout the paper (stationary assumption); however, they are both evolving as the SMBHB merges. The demand that the amplitude of GW-induced white noise  be stationary during the TOA time span set an upper limit on the GW frequency obtained from this method. We estimate the upper frequency limit as follows.
The  GW strain is related to its frequency by \cite{LKJ11}
\begin{equation}
h=C\omega^{2/3},\label{eleven}
\end{equation}
where $C$ is a constant scaling factor determined by the chirp mass ($M_{\rm{chirp}}$) and distance of the GW source.
The GW frequency at the observer is related to the time before final coalescence of the binary $t_{\rm{m}}$ by \cite{Hughes}
\begin{equation}
\omega=2\left(\frac{5}{256}\right)^{3/8}\frac{1}{M_{\rm{chirp}}^{5/8}t^{3/8}_{\rm{m}}(1+z)^{5/8}},\label{twelve}
\end{equation}
where $z$ is the redshift of the GW source ($G=c=1$).
We want the relative change of $K$ in Equation (\ref{five}) to be less 10\% during the time span $\Delta t$ (so that the change of the variance of the GW-induced timing residuals $<1\%$). From Equations (\ref{eleven}) and (\ref{twelve}) we know that
\begin{equation}
1-\frac{\Delta K}{K}=\left(\frac{t_{\rm{m}}-\Delta t}{t_{\rm{m}}}\right)^{1/8}.
\end{equation}
Therefore,
\begin{equation}
\frac{1}{8}\frac{\Delta t}{t_{\rm{m}}}=\frac{\Delta K}{K}<10\%.
\end{equation}
Using Equation (\ref{twelve}) we get the upper frequency limit
\begin{equation}
\omega<2\times\left(\frac{5}{256}\right)^{3/8}(80\%)^{3/8}\Delta t^{-3/8}M_{\rm{chirp}}^{-5/8}(1+z)^{-5/8}.\label{upper}
\end{equation}
Inserting $M_{\rm{chirp}}=1\times10^{8}M_{\odot}$, $\Delta t=1$ yr, and $z<1$ into Equation (\ref{upper}) gives the upper frequency limit of $f_{\rm{up}}=\omega/2\pi\sim2\times10^{-6}$ Hz. Therefore, this method increases the GW frequency upper limit by an order of magnitude without increasing the cadence of observations. Longer $\Delta t$ decreases $f_{\rm{up}}$; however, we can divide the whole data span into small segments and apply the method on each segment. If the chirp mass increases to $M_{\rm{chirp}}=1\times10^{9}M_{\odot}$ then $f_{\rm {up}}\sim5\times10^{-7}$ Hz, which is only a small extension toward the high-frequency end of the detectable GW range by traditional pulsar timing methods.
The impact of relaxing the stationary-amplitude  condition will be studied in the future.
\subsubsection{\textbf{Pulsar term}}
When the GW passes the pulsar, a sinusoidal structure similar to that in Equations (\ref{first}) and (\ref{amplitute}) will be left in the TOA, which is known as the pulsar term. We denote the frequency of the pulsar term and the Earth term as $\nu_{\rm{p}}$ and $\nu_{\rm{E}}$, respectively. $\nu_{\rm{p}}$ can be related to $\nu_{\rm{E}}$ by
\begin{equation}
\nu_{\rm{p}}=\nu_{\rm{E}}\Big(\frac{d(1-\cos\theta)}{ct_m}+1\Big)^{-8/3},
\end{equation}
where $d$ is the distance of the pulsar, and $t_m$ is the time to coalescence, and $\theta$ is the angle between the pulsar and the GW source.
When $\nu_{\rm{p}}<f_{\rm{Ny}}$, we can use a high-pass filter to remove the contribution of the pulsar term and then process as described above.
When $\nu_{\rm{p}}>f_{\rm{Ny}}$, the power contributed by the pulsar term cannot be separated from the Earth term. We denote the amplitude of the signal of the pulsar term and the Earth term as $A_{\rm{p}}$ and $A_{\rm{E}}$, respectively, which are related by
\begin{equation}
\eta\equiv\frac{A^2_{\rm {p}}}{A^2_{\rm {E}}}=\Big(\frac{\nu_{\rm {p}}}{\nu_{\rm {E}}}\Big)^{-2/3}.
\end{equation}
The total rms contributed by both the pulsar term and the Earth term can also be written as
\begin{equation}
\sigma^2_{\rm {GW}}=\frac{1}{2}\mu^2K^2\xi^2\label{useful2}.
\end{equation}
However, in Equation (\ref{useful2}) $\mu^2$ is defined differently compared with that in Equation (\ref{useful}):
\begin{equation}
\mu^2=(1+\eta)(F_{\times}^2+F_{+}^2).\label{redefine}
\end{equation}
When we try to detect the GW source with expected frequency $\nu_{\rm {E}}$, $z,$ and $M_{\rm {Chirp}}$, we can calculate its $t_m$ value via Equation (\ref{twelve}), and then get $\nu_{\rm {p}}$. For each pulsar's $\nu_{\rm {p}}$, we compare it with $f_{\rm{Ny}}$. If $\nu_{\rm {p}}<f_{\rm{Ny}}$, we apply a high-pass filter to the timing residuals and thus remove the pulsar term. If $\nu_{\rm {p}}>f_{\rm{Ny}}$, we refine $\mu^2$ by Equation (\ref{redefine}) and process the data as described above. In the second case, the uncertainty of the distance of the pulsar will affect the determination of $\mu^2$ and thus distort the expected linear correlation in Equation (\ref{useful2}).
Since we need only the variance of white noise, we use the whitened timing residuals. Therefore a large number of pulsars that are not usable in traditional pulsar timing methods because of their red noise can be included in our treatment, including some normal pulsars that have low noise at super-Nyquist Fourier frequencies. With more pulsars and a wider distribution on the celestial sphere, we have a larger range of $\mu^2_i$ and therefore a benefit to GW detection using this method. However, the diversity of the intrinsic white noise level of pulsars also decreases the sensitivity to GW signals.
\vspace*{2mm} \Acknowledgements{\bahao SNZ acknowledges partial funding support from the National Basic Research Program (``973'' Program)
of China (Grant Nos. 2014CB845802 and 2012CB821801), the National Natural Science
Foundation of China (Grant Nos. 11103019, 11133002, 11103022, and 11373036), the Qianren start-up grant 292012312D1117210, and the Strategic Priority Research Program
``The Emergence of Cosmological Structures'' (Grant No. XDB09000000) of the Chinese Academy
of Sciences.}

\end{multicols}
\end{document}